\documentclass{emulateapj}

\usepackage{ifthen}
\newboolean{color}

\setboolean{color}{true}

\def \th{\thinspace}

\def \eg{{{\it e.g.},\ }}

\def \ie{{{\it i.e.}\ }}

\def \viz{{\it viz.\ }}
\def \vs{{\it vs.\ }}
\def \Teff{{$T_{\rm {ef\!f}} $}}
\def \teff{{T_{\rm {ef\!f}} }}
\def\Lo{{$L_\odot $}}
\def\Mo{{$M_\odot $}}
\def\Zo{{$Z_\odot $}}

\def \aanda{A\&A}
\def \Log{{\rm log}}
\def\approxgt{\,\raise2pt \hbox{$>$}\kern-8pt\lower2.pt\hbox{$\sim$}\,}
\def\approxlt{\,\raise2pt \hbox{$<$}\kern-8pt\lower2.pt\hbox{$\sim$}\,}

\long\def\jumpover#1{{}}

\slugcomment{Submitted to ApJ}

\begin{document}

\title{Beat Cepheids as Probes of Stellar and Galactic Metallicity:\\ 
    II. Opacities with the AGS Mixture}


\author{J.~Robert Buchler\altaffilmark{1}}
\altaffiltext{1}{Physics Department, University of Florida,
Gainesville, FL 32611, USA}

\begin{abstract} 

It is well known that the mere location of a Beat Cepheid model in a Period
Ratio \vs Period diagram (Petersen diagram) puts constraints on its metallicity
$Z$.  But these bounds are sensitive to the mixture of elements that are lumped
into the parameter $Z$.  In this short paper we update the previous results
that were based on the Grevesse-Noels solar mixture to the recent, revised
\cite{ags05} (AGS) solar mixture.\\ 
We also examine the effect of the envelope depth on the accuracy of the
computed pulsation periods.  We find that for low period Cepheids with high $Z$
the customary approximation of envelope pulsation breaks down.  It is necessary
to compute stellar models that extend to the center and to include burning and
composition inhomogeneities in the modeling.  Fortunately, however, most Beat
Cepheids that have been observed so far seem to avoid that regime.

\end{abstract}





\maketitle

\section{Introduction} \label{sec:intro}

A recent paper by \cite{BS07} (Paper~I, hereafter) described the results of a
modeling survey of Beat Cepheids that pulsate in the fundamental (F) mode and
the first overtone (O1) simultaneously.  In order to allow a convenient
comparison with the observations, Paper~I presented the results in a $P_{10}$
vs. $P_0$ diagram, a so called Petersen diagram, as a function of metallicity
$Z$, where $P_0$ and $P_1$ are the fundamental and first overtone periods,
respectively, and $P_{10} = P_1/P_0$.  The two periods can easily be extracted
from the observed light curves.  From the metallicity dependent Petersen
diagram one can set bounds on the metallicity $Z$ of the star.  This way of
determining the metallicity is independent of and complementary to all other
stellar and galactic metallicity probes.

\cite{BS07} found that the Petersen diagrams are rather insensitive to stellar
rotation rates and to the galactic helium enrichment function $Y = Y(Z)$ that
they used.  However, the results exhibited sensitivity to the chemical mixture
that is lumped into the metallicity parameter $Z$, mainly through their effect
on the opacities.  The survey of paper I used the OPAL GN93 opacities
(\cite{IR96} for the solar mixture of \cite{GN93}, merged with the \cite{AF94}
opacities at the lower temperatures (log $T<$3.95).

In the last couple of years \cite{ags05} have dramatically revised the solar
chemical composition of metals.  The most important change is a decrease of
elements with atomic numbers $Z<11$, but dominated by O, in favor of elements
with $Z>11$ of which Fe has the most important effect on the opacities.  In
view of the sensitivity of the Petersen diagrams to the chemical mixture we
have thought it useful to update our survey with OPAL opacities that reflect
these new abundances.

In this study, as in Paper I, we use a solar elemental mixture even in low
metallicity Cepheids out of expediency, because there is no observationally
derived information available that would let one improve much upon such an
approximation at this time.

In a linear study such as this the Beat Cepheids are defined as being
simultaneously unstable in the fundamental (F) and the first overtone (O1)
modes.  However, it is well known from nonlinear Cepheid studies (\eg
\cite{kbsc02,kbby98}) that not all stars in this 'linear Beat Cepheid
instability strip' will actually undergo stable Beat pulsations (a large
fraction will be either F {\sl or} O1 Cepheids).  This computationally
convenient, linear definition therefore considerably overestimates the 'Beat
Cepheid instability strip' (broader
by at least a factor of 3 -- 4 in \Log \Teff\ in a
theoretical HR diagram \citep{sb08}.  This in turn increases considerably the
separation of the upper and lower boundaries in the Petersen diagrams with a
concomitant increase in the uncertainty of the stellar metallicities.  The
linear properties of stars do not allow one to infer the modal selection which
is intrinsically nonlinear \citep{stellingwerf74,bk86}.  Therefore, for a
tighter, and correct, definition and study of Beat Cepheids it is necessary to
resort to a full amplitude hydrodynamic survey of models which requires a huge
computing effort.  Such a survey will be presented in a separate paper
(\cite{sb08}).

\begin{figure}
\vskip 30pt
\epsscale{1.1}
\ifthenelse{\boolean{color}}
{\plotone{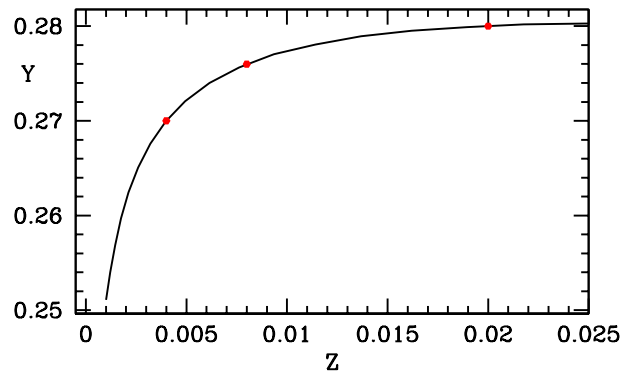}}{\plotone{f1.eps}}
\caption{\small 
Helium galactic enrichment function $Y(Z)$ that is used 
in the model computations.
}
\label{figyz}
\end{figure}

\begin{figure*} 
\epsscale{0.9} 
\ifthenelse{\boolean{color}}
{\plotone{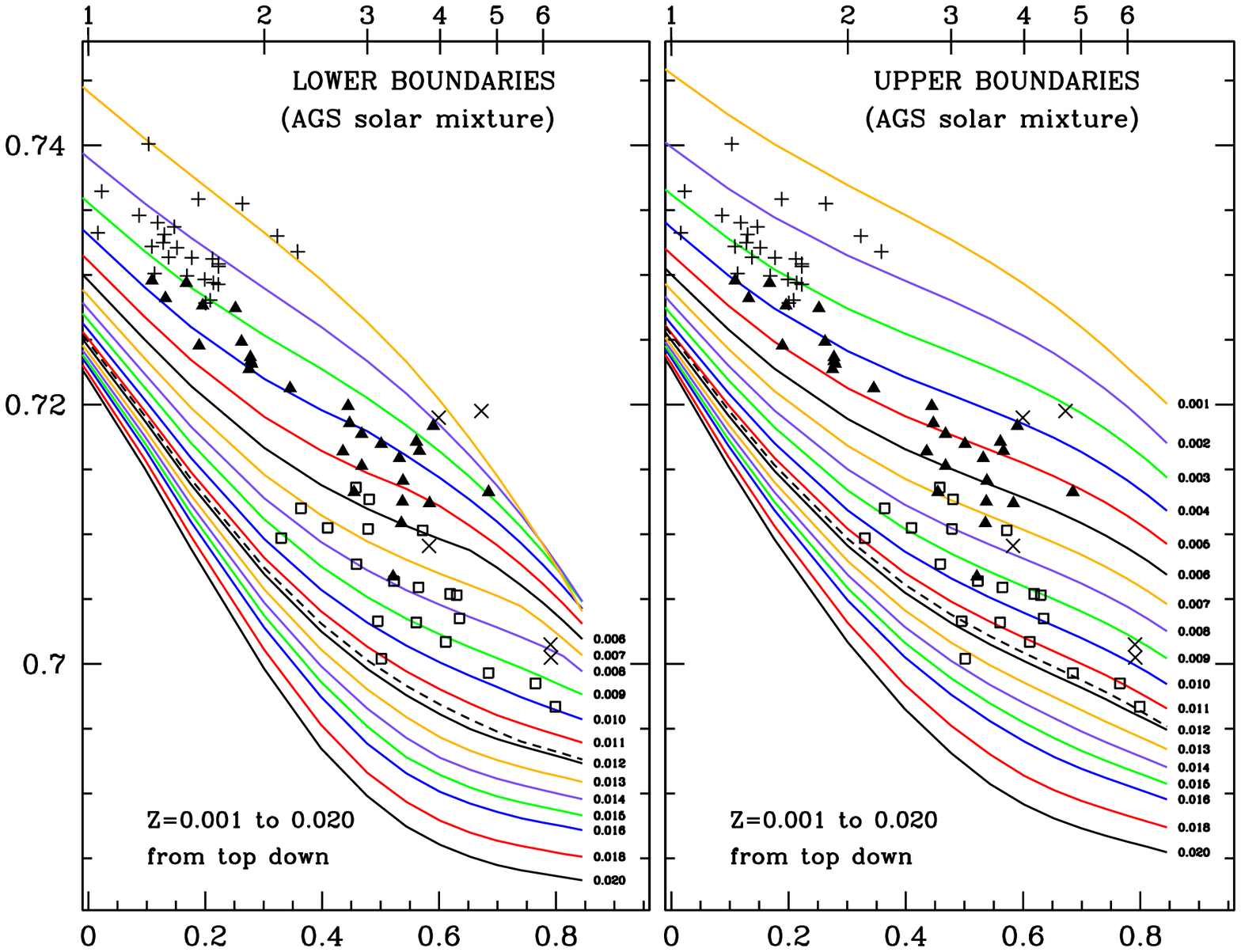}}{\plotone{f2a.eps}}
\vskip 10pt \ifthenelse{\boolean{color}}
{\plotone{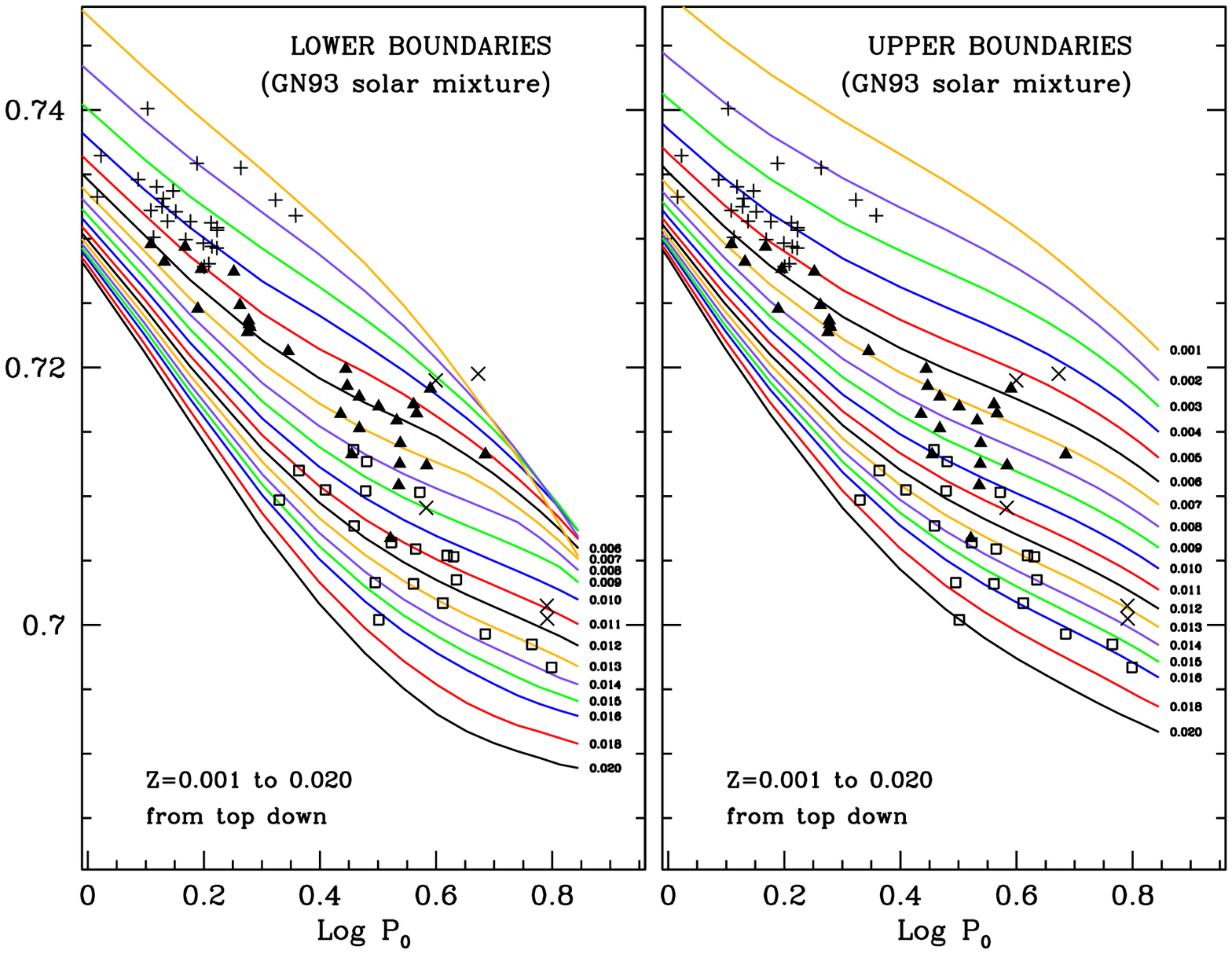}}{\plotone{f2b.eps}}

\caption{\small $P_1/P_0$ \vs \Log $P_0$ \th ($P_0$ on top axis) where $P_0$
and $P_1$ are the fundamental and first overtone periods.  The lines delimit
the ranges for which both F and O1 are linearly unstable: Lower boundaries in
the left, and upper boundaries in the right panels.  The metallicity
increases downward in the figure from $Z$ = 0.001 (top line) to 0.016 in steps
of 0.001, and then from in steps of 0.002 to 0.020 (bottom line).  (In the left
panels the lowest $Z$ values are not shown to avoid overcrowding).  The
long-dashed line refers to the specific AGS composition of the Sun, 
$Z_\odot=0.0122$
and $Y_\odot =0.2486$.  
For reference we have superposed the location of most known
Beat Cepheids.}

\label{figags} 
\end{figure*}

In \S2 we display the computed Petersen diagrams for the AGS solar elemental
composition.  In \S3 we discuss the adequacy of envelope models for Cepheids,
and find that it is necessary to compute full stellar models for the short
period, low metallicity Cepheids.  We conclude in \S4.

\section{Petersen Diagrams with the AGS chemical composition}

This survey proceeds in parallel with the one presented by \cite{BS07} for the
GN93 solar chemical mixture, and we do not repeat a description of the
procedure and of the modeling details here.  We just mention that we have kept
the same helium enrichment function $Y(Z)$ (Fig~\ref{figyz}) that was based on
an interpolation between the 'standard' values for the Galaxy and the two
Magellanic Clouds.  We stress again that the results are very broadly
insensitive to this relation.  We have used the OPAL website to generate
opacities as a function of metallicity $Z$ with the mixture of elements that
are taken from the AGS table of solar abundances.

Our newly computed Petersen diagrams for the AGS mixture are displayed in the
top panels of Fig.~\ref{figags}.  The left-side and right-side panels
show the lower and upper bounds, respectively, of the allowed Beat Cepheid
regions as a function of metallicity $Z$.  We alternate the line-type (or
color) for the successive $Z$ values to make the diagrams easier to use.  In
order to avoid crowding we have omitted the labels 0.001 to 0.005 on the upper
curves in the left-side panels.


\begin{table}
\caption{\small Galactic Metallicities derived from the Beat Cepheids
for the AGS and GN93 elemental mixtures.}
$Z_{min}$ and $Z_{max}$ represent the lowest and largest $Z$ values
that are obtained for all Beat Cepheids in the respective galaxy. 
 $\langle Z\rangle$ is the average of the centroids of the mininum and 
the maximum $Z$ of each star in the galaxy.
\begin{center}
\begin{tabular}{l c c c c c }
\hline\hline
    \noalign{\smallskip}
galaxy  \phantom{donttasemebro} & $\langle Z\rangle$\phantom{toon} 
              & \ & \ \ $Z_{\rm min}$ \ \ & \ & \ \  $Z_{\rm max}$ \ \ \\
    \noalign{\smallskip}
\hline
\hline
    \noalign{\smallskip}
\noalign{\ \ AGS elemental mixture.}
    \noalign{\smallskip}
\hline
    \noalign{\smallskip}
Galaxy & 0.0088 & & 0.0055 & & 0.0133 \\
M33    & 0.0056 & & 0.0000 & & 0.0097 \\
LMC    & 0.0046 & & 0.0024 & & 0.0093 \\
SMC    & 0.0027 & & 0.0010 & & 0.0039 \\
    \noalign{\smallskip}
\hline
    \noalign{\smallskip}
\noalign{\ \ GN93 elemental mixture.}
    \noalign{\smallskip}
\hline
    \noalign{\smallskip}
Galaxy & 0.0118 & & 0.0073 & & 0.0182 \\
M33    & 0.0075 & & 0.0008 & & 0.0124 \\
LMC    & 0.0062 & & 0.0035 & & 0.0124 \\
SMC    & 0.0040 & & 0.0016 & & 0.0069 \\
    \noalign{\smallskip}
\hline
\hline
\end{tabular}
\label{tab1}
\end{center}
\end{table}

The dashed curves in the top panels in Fig.~\ref{figags} show the results for
the opacity with the specific AGS composition for the Sun, \viz
$Z_\odot=0.0122$ and $Y_\odot =0.2486$.  Despite the large difference in $Y$
(this point lies below the range of the diagram in Fig.~\ref{figyz}),
this curve falls close to the $0.012$ lines, illustrating again the
insensitivity to the exact values of $Y$.

For an easy comparison we have also juxtaposed the GN93 composite Petersen
diagram of Paper I in the bottom panels of Fig.~\ref{figags}.  The GN93 models
have been recomputed with the same zoning (200 mass-shells) as for the new AGS
results, which is a little finer than that used in \cite{BS07}.  It is
noteworthy that the AGS mixture causes an overall downward shift of the curves,
especially for the shorter periods. and a downward stretching for the longer
periods.  One therefore obtains generally lower metallicities with the AGS
solar mixture than with the GN93 solar mixture.  This is of course expected
since AGS have revised the metallicity for the Sun down to $Z_\odot=0.0122$.

The locations of the known Beat Cepheids are superposed in Fig.~\ref{figags} as in
\cite{BS07}, with open squares for the Galaxy, x's for M33, filled
triangles for the LMC and crosses for the SMC.  We have limited the figure to
periods longer than one day despite the fact that a few such stars are known in
the SMC.  The reason will be given in \S\ref{sec:disc}.

The diagrams of Fig.~\ref{figags} are used as follows: If a Beat Cepheid falls
between the upper and lower boundaries for a given $Z$ value, that means that
models with that value of $Z$ are compatible with the observed period
constraints for the given Beat Cepheid.  By finding all the compatible $Z$
values one can set upper and lower bounds on $Z$.  In practice, one
interpolates between the displayed curves in order to refine the $Z$ bounds.

In Table~\ref{tab1} we present the metallicities for the 4 galaxies in which a
significant number of Beat Cepheids are known.  From the composite Petersen
diagram we infer for each star (i) the range of possible $Z$, \ie $Z_{\rm min,
i}$ and $Z_{\rm max, i}$ which, as we have seen, represent the uncertainty
range of $Z$ for each Beat Cepheid (due to the width of the instability strip).
We define $Z_{\rm min}$ as the minimum over the galactic sample of these
$Z_{\rm min, i}$ and $Z_{\rm max}$ as the maximum of the $Z_{\rm max, i}$.
Finally we introduce the galactic average $\langle Z\rangle$ as the average of
the centroids of the ranges $\frac12 (Z_{\rm min,i}+Z_{\rm max,i})$.

With the new opacities we obtain an average metallicity of $\langle Z\rangle$ =
0.0088 and a range of uncertainty of 0.0055 to 0.0133 for the Galactic Beat
Cepheids.  This is to be compared with an average $\langle Z\rangle$ = 0.0118
and a range of 0.0073 to 0.0182 with the GN93 opacities.  As noted in Paper~I
the GN93 average $\langle Z\rangle$ from the Galactic Beat Cepheids is quite
low compared to the GN93 value for the Sun \Zo $= 0.017$, the but the AGS the
average $\langle Z\rangle$ of the Beat Cepheids is now much closer to the AGS
value of \Zo=0.0122.

We note again here that the large, indicated spread $Z_{\rm min}$ to $Z_{\rm
max}$ represents a combination of the actual spread in the galactic
metallicity, and of the uncertainty that arises from the use of the linear
definition of Beat Cepheids.  As mentioned earlier, we hope to reduce the
latter with a survey of full amplitude Beat Cepheid models \cite{sb08}.

\begin{figure}
\epsscale{1.0}
\ifthenelse{\boolean{color}}
{\plotone{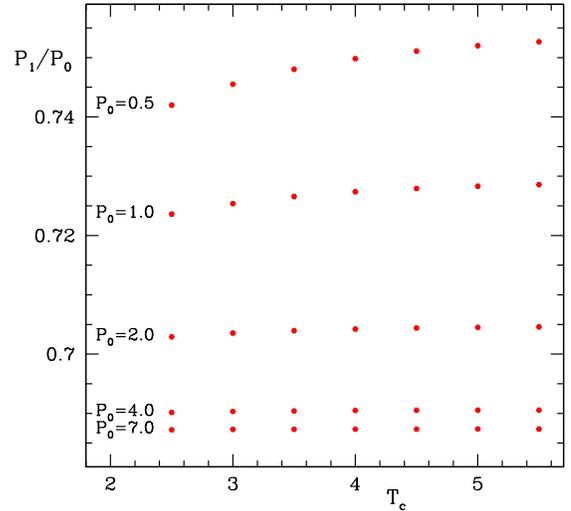}}{\plotone{f3.eps}}
\caption{\small 
The period ratio $P_{10}$ as a function of the temperature of the 
envelope depth, $T_c$ (in 10$^{6}$K), 
 for the sequences of models of Table~1 with the indicated fundamental periods
(in days).}
\vskip 20pt
\label{figp10depth}
\end{figure}

\begin{figure}
\epsscale{1.15}
\ifthenelse{\boolean{color}}
{\plotone{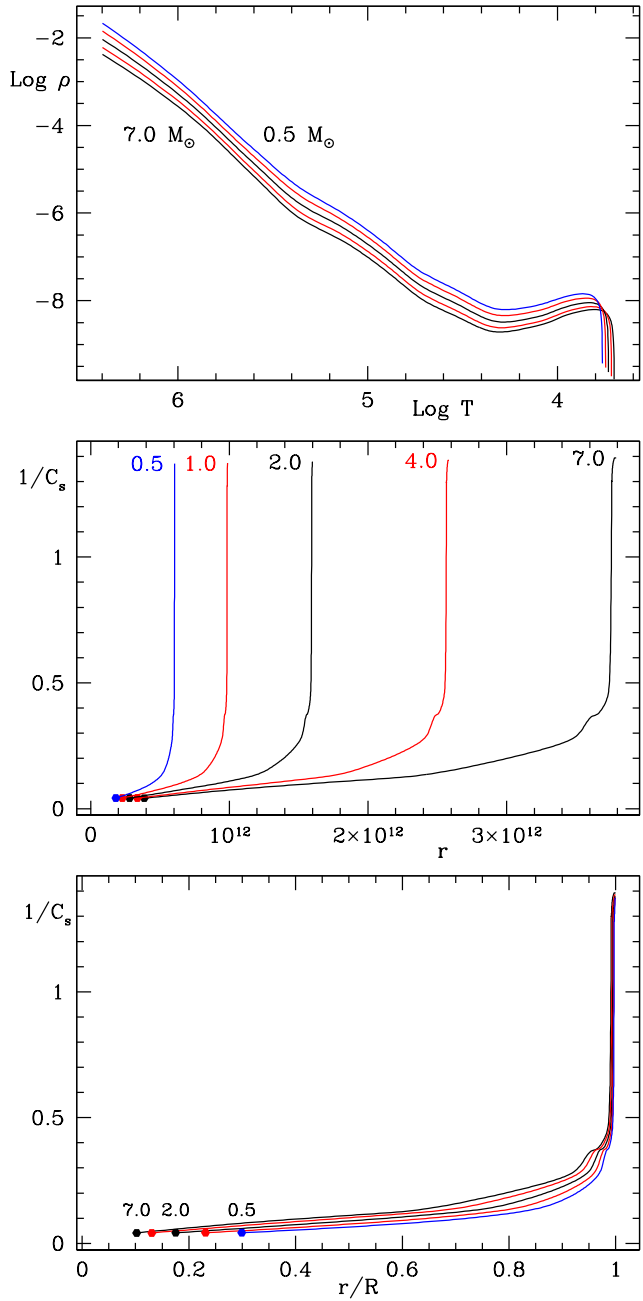}}{\plotone{f4.eps}}
\caption{\small 
(a) Model structure, \Log \th $\rho$\th[g/cm$^3$] \th vs. 
     \Log\th T\th [K] \th for the 5 models 
     of Table~2 on the fundamental blue edge
   with fundamental periods, 0.5, 1.0, 2.0, 4.0 and 7.0\thinspace d;
(b) Inverse sound speed (in units of $10^6$ cm/s) vs. radius [cm];
(c) Inverse sound speed vs. relative radius.
}
\label{figplotall}
\end{figure}

\begin{figure*}
\begin{center}
\epsscale{1.0}
\ifthenelse{\boolean{color}}
{\plotone{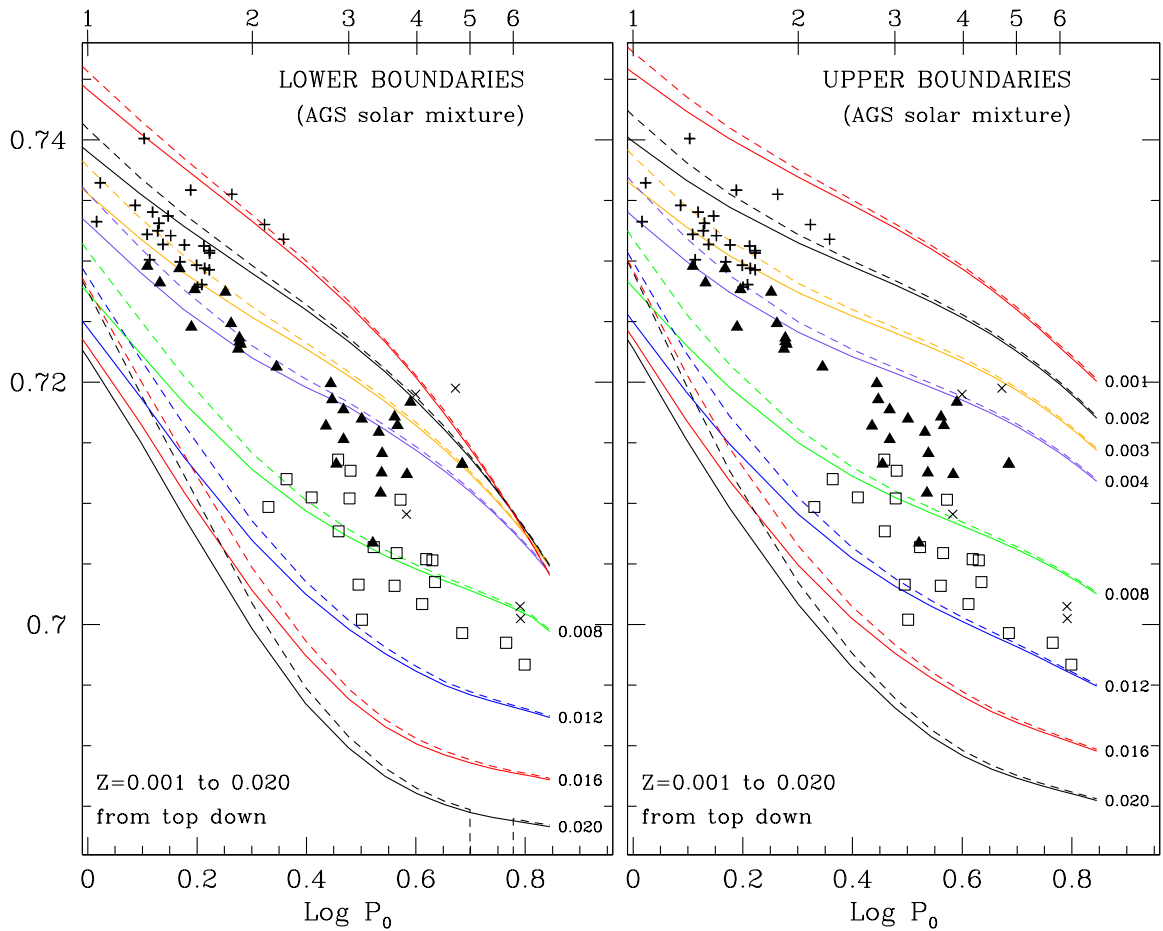}}{\plotone{f5.eps}}
\vskip 10pt
\caption{\small 
\small Sensitivity to the depth of the envelope $T_c$ of the
$P_1/P_0$ \vs \Log $P_0\thinspace$ [d] Petersen diagrams\th 
(for an AGS mixture). 
The solid lines show the results for an envelope of depth 
 $T_c$ = 2.5 $\times 10^6$\th K
and the dashed curves for $T_c$= 4.5 $\times 10^6$\th K. 
 Same notation as in Fig.~\ref{figags}.
}
\vskip 20pt
\label{figcomp200depth}
\end{center}
\end{figure*}

\vskip 20pt

\section{Discussion of the Modeling Uncertainties\\
-- Inadequacy of Envelope Models}\label{sec:disc}

Ever since the early days of Cepheid modeling, it has been customary to
approximate Cepheids with envelope models of uniform composition, because the
pulsation is largely confined to the outer parts of the stars.  One thus
imposes a rigid core of radius $R_c$ at some temperature $T_c$, located outside
the burning region where the luminosity $L_c$ is constant.  This approximation
has been justified on the one hand, because the radial displacement
eigenvectors for the low lying pulsation modes decay rapidly inward and one
therefore sets $\delta r=0$ at $r=R_c$, and, on the other hand, because the
computed periods are found not to change much when one varies the depth of the
envelope models.  However, for the Petersen diagrams we require a higher
relative accuracy ($<0.001$) in the periods, as we now discuss.

\begin{table}
\caption{\small Model Sequences used in the Tests of \S3}
\begin{center}
\begin{tabular}{r r r r r r}
\hline\hline
    \noalign{\smallskip}
\ \ \ $P_0$ [d] & \ \ $M$/\Mo \ \  & \ \ $L$/\Lo\ \ & \ \ $\teff$ [K] \ \  & \
\  $R_*$[10$^{11}$ cm] \\
    \noalign{\smallskip}
\hline
\hline
  0.5 & 3.09&   154& 6882&   6.0821 \\
  1.0 & 3.63&   347& 6612&   9.8743 \\
  2.0 & 4.32&   805& 6407&  16.0227 \\
  4.0 & 5.14&  1806& 6175&  25.8364 \\
  7.0 & 5.93&  3422& 5982&  37.8964 \\
    \noalign{\smallskip}
\hline
\hline
\end{tabular}
\label{tab2}
\end{center}
\end{table}

A reexamination of this assumption shows a breakdown for models with low
periods and high $Z$.  It turns out that in this regime the periods, and
concomitantly the period ratios, exhibit an uncomfortable dependence on $R_c$
or $T_c$.  Furthermore, for the lowest periods there is a poor leveling off
even with $T_c$ up to (perhaps unreasonably large) values of 5.5$\times
10^6$\th K which is beyond the temperature up to which one can neglect nuclear
burning and spatial composition inhomogeneities.

This $T_c$ dependence is illustrated in Fig.~\ref{figp10depth}, which plots the
period ratios $P_{10}$ \vs $T_c$ for a sequence of 5 models (see
Table~\ref{tab2}) that have been chosen with a composition of $X=0.7045$ and
$Z=0.016$ and that are all located at the high temperature side of the region
where both the fundamental and the first overtone modes are linearly unstable.
The fundamental periods of the sequences are $P_0$ = 0.5, 1.0, 2.0, 4.0, and
7.0\thinspace d, respectively.  The shape for the leveling off is found to be
the same for all sequences, but while the variation in $P_{10}$ remains within
an acceptable range for the long period models, this is not the case for the
short period models.  For example, in the 4\thinspace d sequence a change from
$T_c$ = 2.5 $\times 10^{6}$ to 3.0 $\times 10^{6}$\th K leads to an increase of
$P_{10}$ of less than 0.0002, but for the 0.5\thinspace d sequence the same
change results in an increase of $P_{10}$ of 0.0035.  This is a large amount if
one considers the Petersen diagram of Fig.~\ref{figags}.  Furthermore, for the
low period models the leveling off is not complete even by the time one
reaches $T_c = 5.5\times 10^6$\th K.

In Fig.~\ref{figplotall} we present some other relevant properties of the
sequence of models from Table~\ref{tab2}.  Thus the top panel displays the
models in the $\Log\th\rho$ \vs $\Log\th T$ plane, and reveals a very similar
structure for the 5 models and demonstrates that nothing special is occurring
for the 0.5\thinspace d models.

The middle (b) and bottom (c) panels exhibit the inverse of the sound speed \vs
radius, $r$, and \vs relative radius ($r/R_*$), respectively, where $R_*$ is
the equilibrium stellar radius.  The dots correspond to the location of the
rigid inner core $R_c$, defined to lie at $T_c =2.5 \times 10^6$\th K in
these models.  In actual radius, the low period models are much deeper, but
this is not the case in relative radius.  The reason for presenting these
quantities is that the fundamental period $P_0$ is roughly proportional to the
stellar sound traversal time $$P_0 \propto \int {1\over c_s} dr.$$ \noindent
The combined effect of the high sound speed in the stellar core and the decay
to zero of the eigenvectors cause the contribution from the neglected
penetration of the core to be very small in general.  It is clear from
Fig.~\ref{figplotall}(b) that the contribution to the period from a deepening
of the envelope by $\Delta R_c$, namely $\Delta R_c / \langle C_s\rangle$, is
larger for the long period models.  However, when one considers it relative to
the period, one finds a much larger contribution for the envelope models with
the smaller periods.  For example, from the figure we infer that in the
7.0\thinspace d model a small change of 10$^{10}$\th cm in $R_c$, \ie $\Delta
R_c/R_*\sim0.003$, would cause an increase of $P_0$ by $\sim 0.005$\thinspace
d, \ie a $\sim$ 0.07\th\% relative period change.  In the 0.5\thinspace d model
the same change of $R_c$, \ie $\Delta R_c/R_*\sim0.02$, would cause a
considerable $\sim$1 \% change in the period.

Fortunately, an accuracy in the pulsation periods of a percent or less is
generally obtainable and sufficient for most purposes, so that envelope models
are therefore quite adequate.  Here, however, one puts higher demands on the
Cepheid models, because one wants to extract a Cepheid's metallicity from its
location in the Petersen diagrams, for which a relative accuracy of $\sim
10^{-3}$ is necessary.

In Fig.~\ref{figcomp200depth} we present a composite Petersen diagram that
compares Beat Cepheid models in which the depth is set at our standard
$T_c=2.5\times 10^6$ K (solid lines) with those at $T_c = 4.5\times 10^6$ K
(dashed lines).  One notices that the most unreliable models are to be found in
the high $Z$, low period domain.  Luckily, the vast majority of the observed
Beat Cepheids avoid the regime in which the uncertainties are largest.

Fig.~\ref{figcomp200depth} shows that one would be ill advised to use envelope
models for the $\approxlt 1.0$\th d stars that are found in the SMC
(\cite{mbb07}).  For these Beat Cepheids it is necessary to use models that
extend deeper, if not full stellar models.  This is, of course, a much
harder problem, in that it requires the inclusion of nuclear burning and of the
spatial composition variations that occur during stellar evolution.  In other
words, one needs a stellar evolution code that incorporates a linearization to
compute the fully nonadiabatic pulsation frequencies with a time-dependent
mixing length recipe to capture the pulsation -- convection interaction.
 
\section{Conclusion}

We have computed the Beat Cepheid models that make up the composite Petersen
diagrams as a function of metallicity $Z$ with the \cite{ags05} solar
mixture.  This change causes a noticeable reduction of $\sim$ 25 - 30 \% in
the average $\langle Z\rangle$ of the four galaxies in which a sizable number
of Beat Cepheids are known.  It considerably improves the agreement between the
average Galactic metallicity and the Sun's metallicity over what was obtained
with the GN93 abundances.

While the elemental mixture of the low $Z$ stars is most likely not quite
solar, this is the best approximation we can use in the absence of detailed
determinations of chemical compositions from high resolutoin spectra of such
stars.

Because we needed to construct Beat Cepheid models down to periods as short as
0.5\thinspace d as have been found in the SMC by \cite{mbb07}, we have been led
to reexamine the accuracy of commonly used envelope models for Cepheid stars.
We find that because of the high accuracy demands of Petersen diagrams such
envelope models become inadequate for short period, high $Z$ Cepheids.  Good
fortune has it though that the higher $Z$ Beat Cepheids in the Galaxy and in
M33 have longer periods, and that the Beat Cepheids with smaller periods are
observed in the lower metallicity Magellanic Clouds.  Thus the vast majority of
the known Beat Cepheids avoid to some extent the troublesome region of
uncertainty of envelope models.

\begin{acknowledgements}

We wish to thank Daniel Cordier, Robert Szab\'o and Peter Wood for valuable
discussions.  We are also very grateful to Carlos Iglesias and Forrest Rogers
for making the computation of opacity tables on the OPAL homepage so
convenient.  Finally, we wish to thank an anonymous referee whose comments have
improved this paper.  This work has been supported by NSF (AST07-07972 and
OISE04-17772) at UF.

\end{acknowledgements}


\end{document}